\documentclass[conference]{IEEEtran}
\usepackage{cite}
\usepackage{amsmath,amssymb,amsfonts}
\usepackage{algorithmic}
\usepackage{graphicx}
\usepackage{textcomp}
\usepackage{xcolor}
\def\BibTeX{{\rm B\kern-.05em{\sc i\kern-.025em b}\kern-.08em
    T\kern-.1667em\lower.7ex\hbox{E}\kern-.125emX}}

\newcommand{\etavec}{\mbox{\boldmath$\eta$}}

\newcommand{\wvec}{\mbox{\boldmath$w$}}

\begin{document}

\title{Harnessing adaptive dynamics in neuro-memristive nanowire networks for transfer learning\\
}


%
\author{\IEEEauthorblockN{
Ruomin Zhu\IEEEauthorrefmark{1},
Joel Hochstetter\IEEEauthorrefmark{1},
Alon Loeffler\IEEEauthorrefmark{1},
Adrian Diaz-Alvarez\IEEEauthorrefmark{2}, 
Adam Stieg\IEEEauthorrefmark{2}\IEEEauthorrefmark{3},
James Gimzewski\IEEEauthorrefmark{2}\IEEEauthorrefmark{3},\\
Tomonobu Nakayama\IEEEauthorrefmark{2}\IEEEauthorrefmark{1}
and
Zdenka Kuncic\IEEEauthorrefmark{1}\IEEEauthorrefmark{2}
\IEEEauthorblockA{\IEEEauthorrefmark{1}School of Physics and Sydney Nano Institute,
University of Sydney, Sydney, NSW 2006, Australia\\ Email: zdenka.kuncic@sydney.edu.au}
\IEEEauthorblockA{\IEEEauthorrefmark{2}International Centre for Materials Nanoarchitectonics, National Institute for Materials Science, Tsukuba, Japan}
\IEEEauthorblockA{\IEEEauthorrefmark{3}California NanoSystems Institute, University of California at Los Angeles, California, USA}
}}


\maketitle

\begin{abstract}
Nanowire networks (NWNs) represent a unique hardware platform for neuromorphic information processing. In addition to exhibiting synapse-like resistive switching memory at their cross-point junctions, their self-assembly confers a neural network-like topology to their electrical circuitry, something that is impossible to achieve through conventional top-down fabrication approaches. In addition to their low power requirements, cost effectiveness and efficient interconnects, neuromorphic NWNs are also fault-tolerant and self-healing. These highly attractive properties can be largely attributed to their complex network connectivity, which enables a rich repertoire of adaptive nonlinear dynamics, including edge-of-chaos criticality. Here, we show how the adaptive dynamics intrinsic to neuromorphic NWNs can be harnessed to achieve transfer learning. We demonstrate this through simulations of a reservoir computing implementation in which NWNs perform the well-known benchmarking task of Mackey-Glass (MG) signal forecasting. First we show how NWNs can predict MG signals with arbitrary degrees of unpredictability (i.e. chaos). We then show that NWNs pre-exposed to a MG signal perform better in forecasting than NWNs without prior experience of an MG signal. This type of transfer learning is enabled by the network's collective memory of previous states. Overall, their adaptive signal processing capabilities make neuromorphic NWNs promising candidates for emerging real-time applications in IoT devices in particular, at the far edge.
\end{abstract}


\section{Introduction}

The field of neuromorphic engineering is widely recognized as the realization of Carver Mead's original vision for a new type of electronic hardware engineered to mimic information processing in biological nervous systems \cite{Mead-1990, Mead-2020}. Today, the lowest common denominator of virtually all neuromorphic hardware systems is the co-location of memory and processing units (i.e. non-von Neumann architecture). This minimal neuromorphic feature alone has dramatically improved power efficiency in training various artificial neural network (ANN) models \cite{Zhang-2020}.

A higher level neuromorphic attribute is the ability to learn and while ANN models demonstrate learning in software, learning in hardware is desirable for next-generation stand-alone cognitive devices, especially at the IoT edge \cite{edge}. In hardware, spike-based learning has been successfully implemented in conventional silicon CMOS technology (e.g. \cite{Indiveri-2011,Merolla-2014,Brainscales1,Brainscales2}).
Beyond silicon, nanoelectronic materials with intrinsic neuromorphic properties, including memory and the ability to emulate synaptic connections \cite{Burr-2017,Zhang-2020}, have attracted enormous attention for on-chip learning \cite{Sangwan-2020}. In particular, resistive switching memory (memristive) devices \cite{Waser-Aono-2007,Zidan-2018,Ielmini-Wong-2018,Wang-2020} are leading candidates for efficient neuromorphic computing architectures, with demonstrated neuromorphic learning functionalities such as short-term and long-term potentiation (STP/LTP) and spike-timing dependent plasticity (STDP) \cite{Ohno-2011,Serrano-2013,Serb-2016,Mehonic-2020}. 

At the device level, neuromorphic functionalities can be broadly attributed to modification of electronic transport mechanisms by nanoscale geometric confinement, usually across a metal-insulator-metal (MIM) junction \cite{Pershin-DiVentra-2011}. Importantly, synapse-like memristive switching is observed not just in memristors fabricated from conventional bulk materials (e.g. metal oxides), but also in neuromorphic systems self-assembled from nanomaterials using bottom-up techniques \cite{Sangwan-2020}. Here, we focus on self-assembled metallic nanowires because not only do they form memristive switching MIM junctions, but they also form a complex neural-like network topology, with all-in-one connectivity properties such as small-worldness, modularity and recurrent feedback loops \cite{Nirmalraj-2012,Bellew-2014,Avizienis-2012,Demis-2015,Milano-2019,Diaz-Alvarez-2019,Pantone-2018,Loeffler-2020}. The unique neuromorphic topology of self-assembled nanowire networks (NWNs) is responsible for collective functionalities emerging from the interplay between network connectivity and synaptic nonlinear dynamics \cite{Stieg-2012,Sillin-2013,Manning-2018,Kuncic-2018,Diaz-Alvarez-2019,Milano-2020}.

Learning in NWN hardware does not require implementation of an ANN model, as has been demonstrated with associative memory tasks \cite{Diaz-Alvarez-2020,Li-2020} and with temporal information processing tasks using a reservoir computing approach, where the network self-regulates in response to continuous-time input signals and only the readout is trained \cite{Sillin-2013,Fu-2020,Kuncic-2020}.
Varying spatio-temporal input signals (i.e. delivered via different contact electrodes and with time-varying amplitudes) results in the formation of new electrical pathways, analogous to synaptogenetic learning \cite{Zito-2002,Cui-2016}.  
Here, we show that NWNs with prior experience of a complex, nonlinear time-series signal can perform better in forecasting the signal than a NWN without prior exposure, thus demonstrating capacity for transfer learning, an important attribute for general intelligence (see \cite{Zhuang-2020} for a recent comprehensive review).

\section{Methods}

\subsection{Modelling network connectivity and memristive junctions}

\begin{figure}
    \centering
    \includegraphics[width=1.7in]{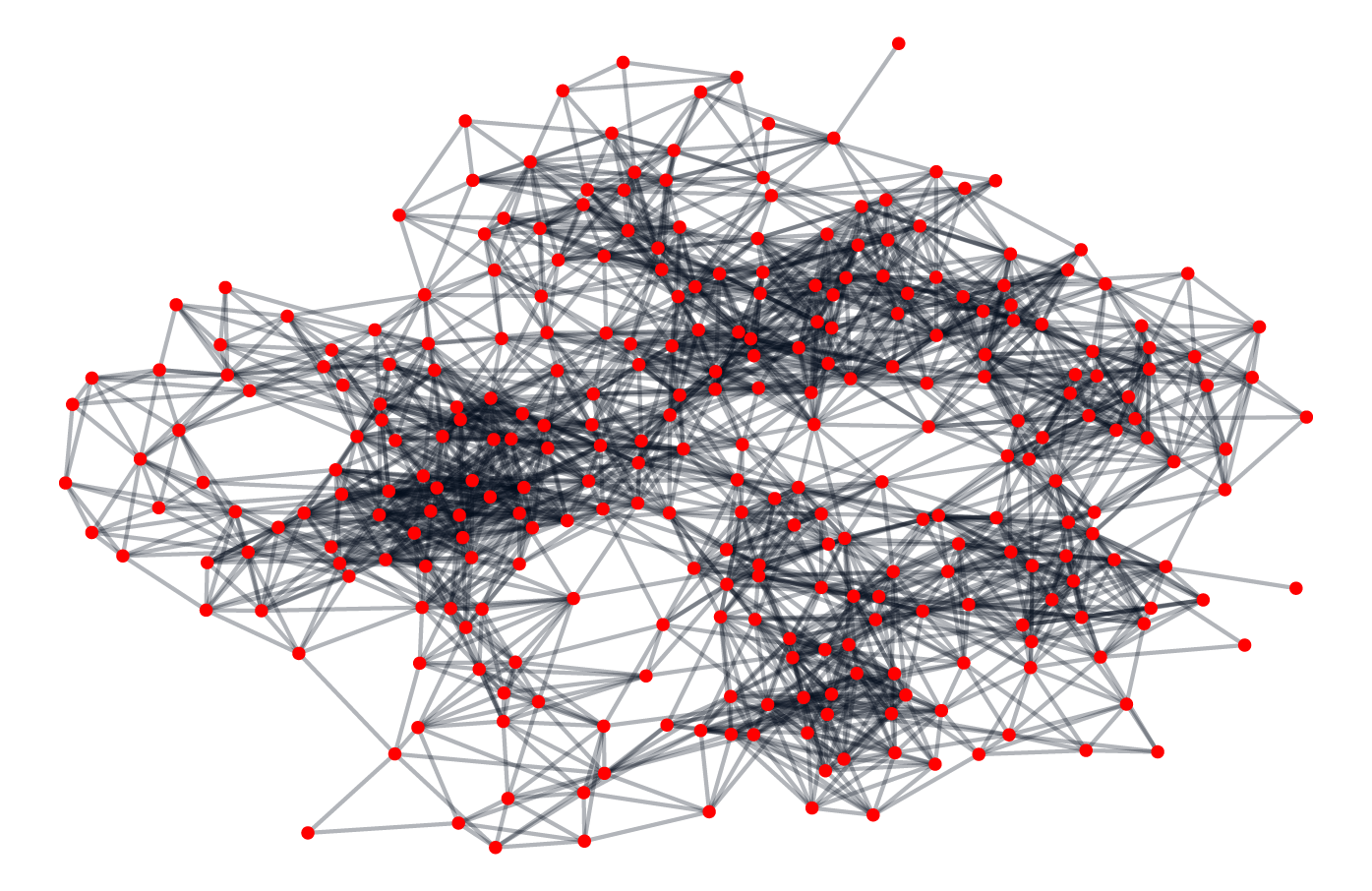}
    \includegraphics[width=1.7in]{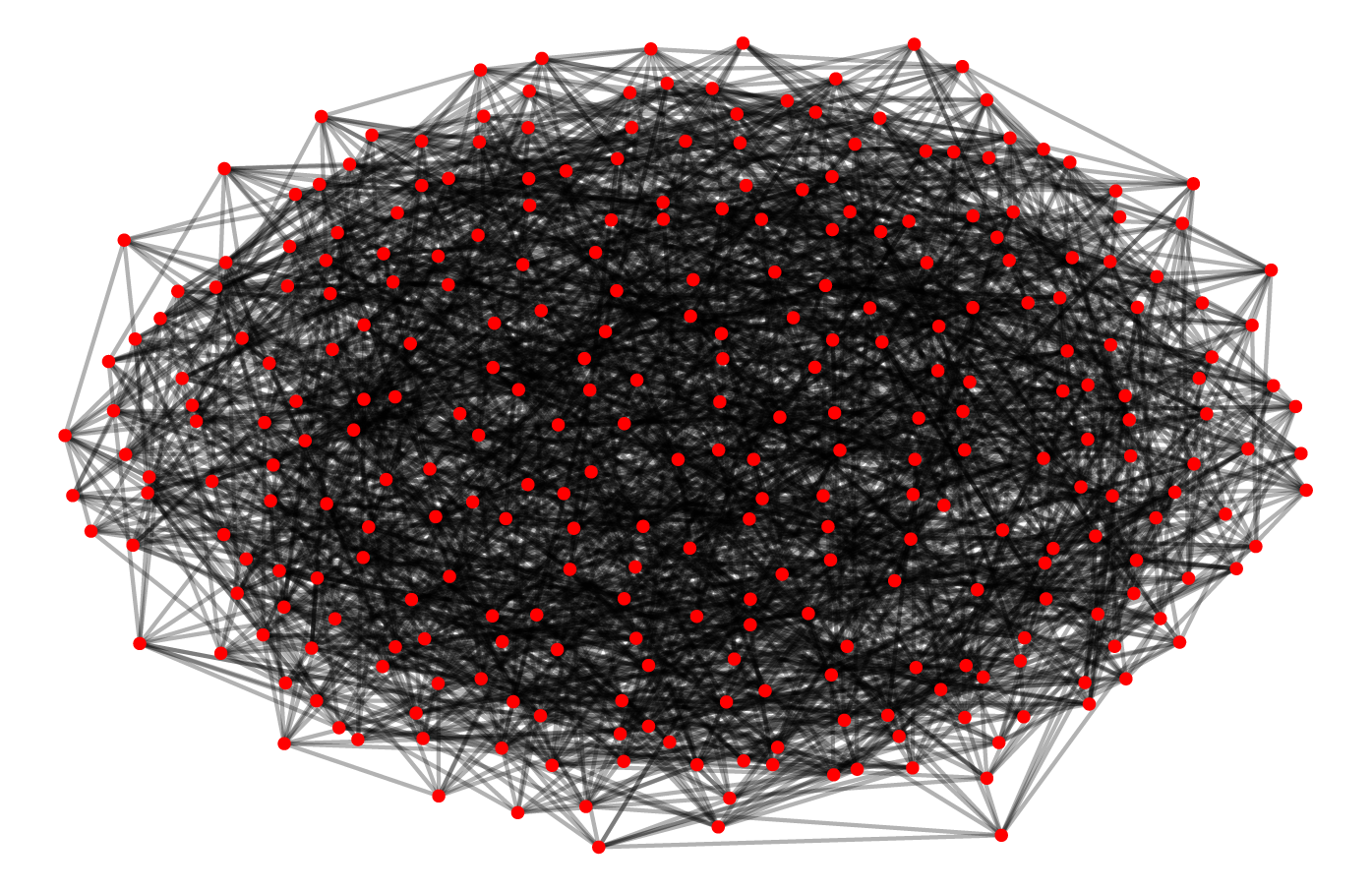}
    \caption{Graph representations of 300-node networks: left -- self--assembled nanowire network (2434 edge junctions, average degree 16, small-world propensity 0.67); right -- random network (2400 junctions, average degree 16, small-world propensity 0.29). Nodes in red, edges in black.}
    \label{fig:networks}
\end{figure}

We performed simulations using a physically motivated model based on polymer-coated Ag nanowires that self--assemble into a complex network \cite{Kuncic-2018,Diaz-Alvarez-2019}. Self--assembly was modelled by distributing individual nanowires on a 2D plane, with uniformly random positions and orientations, and with lengths uniformly sampled from a gamma distribution (mean 100\,$\mu$m, stdev 10\,$\mu$m). The variance in nanowire length is based on experimental observations \cite{Nirmalraj-2012,Bellew-2014,Diaz-Alvarez-2019,Diaz-Alvarez-2020} and increases the probability of forming cross-point junctions between overlapping nanowires. This mimics biological neural networks, in which individual neurons can each make several thousand synaptic connections to neighbouring neurons. In our model of self-assembled nanowire networks (NWNs), a range of nanowire connectivities are possible for a fixed number of nanowires. Importantly, the resulting network structure is more complex than a purely random topology or fully connected network (Fig.~\ref{fig:networks}), with sparseness and recurrence characteristics that are responsible for efficient signal transduction and emergent cognitive function in biological neural networks \cite{Bullmore-Sporns-2009,Lynn-Bassett-2019,Loeffler-2020}. It is also noteworthy that the complex network topology of self-assembled networks differs from the bipartite structure used in ANN models.

Nanowire-nanowire cross-points were modelled as voltage-controlled memristive junctions described by a state-dependent Ohm's law, $I = G (\lambda ) V$, where the conductance $G(\lambda)$ is a function of the state variable $\lambda (t)$ that depends on the past history of voltage input. Physically, $\lambda (t)$ parameterizes the evolution of a conductive filament that forms across the MIM junction above a threshold bias. For polymer-coated Ag nanowires, the polymer is electrically insulating, but ionically conducting, so Ag$^{+}$ cations can migrate across the biased junction \cite{Zhu-2020}. The conducting atomic filament that forms in this way switches the junction from a high-resistance ``off'' state, to a low-resistance ``on'' state, when $\lambda \geq \lambda_{\rm crit}$, where $\lambda_{\rm crit}$ is a threshold. As the polymer thickness ($\approx 2-3\,$nm) is comparable to the Fermi length of Ag ($\approx 0.5\,$nm), resistive switching is modelled as a change in the junction conductance state $G(\lambda)$ by an amount equal to the conductance quanta $G_0 = (13\,{\rm k}\Omega)^{-1}$, consistent with measurements of individual nanowire junctions \cite{Terabe-2005}. The corresponding resistance states are $R_{\rm on} = G_0^{-1}$ and $R_{\rm off} = \zeta R_{\rm on}$, with $\zeta = 10^4$ used in the simulation results presented here. Additionally, the Simmons formula is used to model the low voltage tunneling regime in $G(\lambda)$ when the conductive filament is close to the opposite nanowire \cite{Simmons-1963}. Network conductance is calculated using a modified nodal analysis \cite{Ho-1975}  to solve Kirchoff's circuit law equations at each time point.

\subsection{Mackey--Glass time series prediction}

Reservoir computing was implemented on a network with $N=100$ nanowire nodes and 577 memristive junctions. The Mackey--Glass (MG) signal was delivered to one source node as input voltage bias relative to a drain node. MG signals with varying time delays $\tau \geq 17$ were predicted, with $\tau = 17$ corresponding to the onset of chaotic dynamics.




The network state as a function of time $t$ is represented by instantaneous voltage on all the $N$ nodes. The MG signal at a future time $u_{t+\delta t}$ was predicted using a subset of $n=10$ node states weighted by a vector $\wvec$:
\begin{equation}
    u_{t+\delta t} = \wvec \cdot \etavec_t,
    \label{e:predict}
\end{equation}
where $\etavec_t$ is an 11-element vector that includes a 1\,V linear shift element and where $\wvec$ was determined by least squares regression using all past states of the $n$ nodes and the input (teacher) signal in the time interval $t \in [0,T]$, i.e.
\begin{equation}
    [u_{\delta t}, u_{\delta t+1}, ... , u_{T}]
    = \wvec \cdot [\etavec_0, \etavec_1, ... \etavec_{T - \delta t}].
\end{equation}
A history length of $T > \tau$ was used to train the $n$ output weights and the prediction step was set to $\delta t = \tau$.
Accuracy of the prediction task was calculated as
\begin{equation}
    \mbox{Accuracy} = 1 - \mbox{RNMSE}
\end{equation}
where RNMSE is the root-normalized mean square error.
Statistical uncertainties were determined by randomly selecting the $n=10$ readout nodes for 100 simulations and averaging the accuracy.




\subsection{Transfer learning}

In conventional reservoir computing, the initial state of the network is homogeneous and for the MG prediction task described above, we set $\etavec_0 =$ \boldmath{$0$}. We modified the task by first exposing the network to a source MG signal with delay parameter $\tau_1$ before training and predicting a second target MG signal with delay parameter $\tau_2$. Delivering the source MG ($\tau_1$) signal for 1.5\,s effectively primed the network to an initial state $\etavec_0^\prime \neq$ \boldmath{$0$} that has memory of previous states associated with the source MG signal. This is analogous to transfer learning methods applied to ANN models, where synaptic weights are trained on a source domain and the knowledge gained is transferred to a different, but related, target domain \cite{Zhuang-2020}. Our case is somewhat different as the network dynamically self-adjusts its own synaptic junction states during the priming period (since in reservoir computing, only the output weights are trained, not the network weights).  
We compared the accuracy in predicting the MG target signal to that obtained for a network without prior exposure to the source MG signal during a pre-training period.

\section{Results}

\subsection{Adaptive dynamics}

\begin{figure}[h]
    \centering
    \includegraphics[width=3.3in]{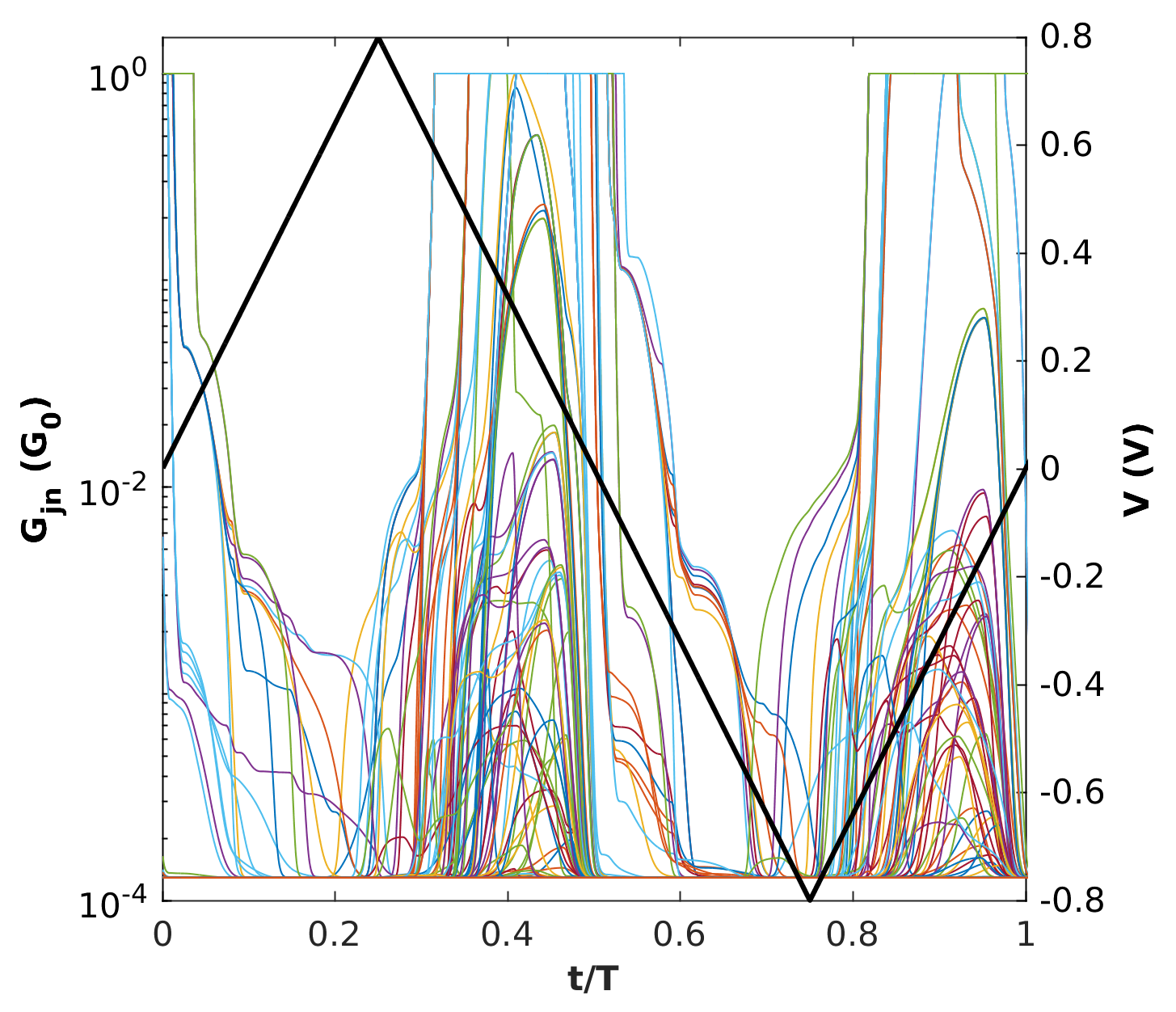}
    \includegraphics[width=1.7in]{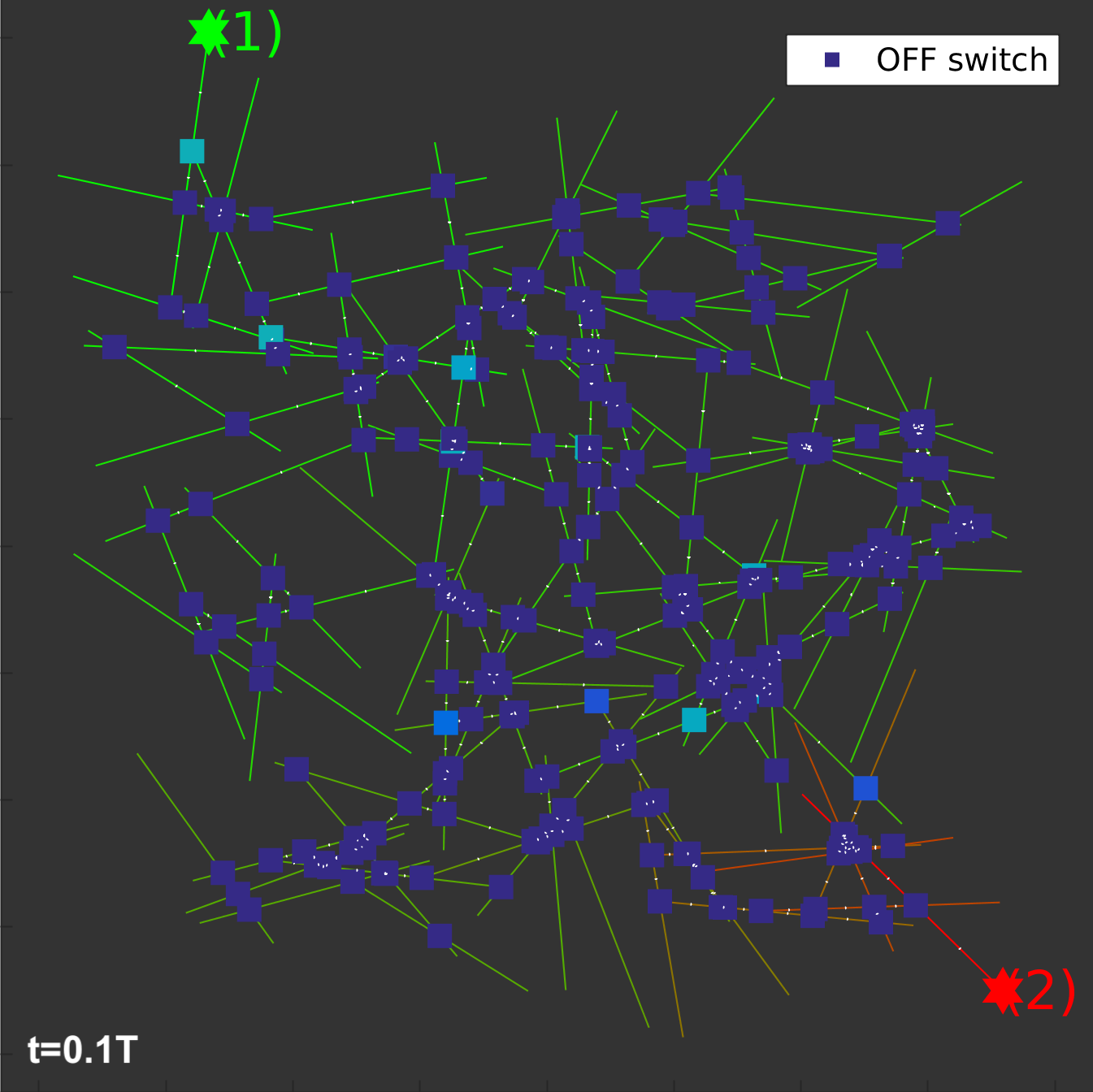}
    \includegraphics[width=1.7in]{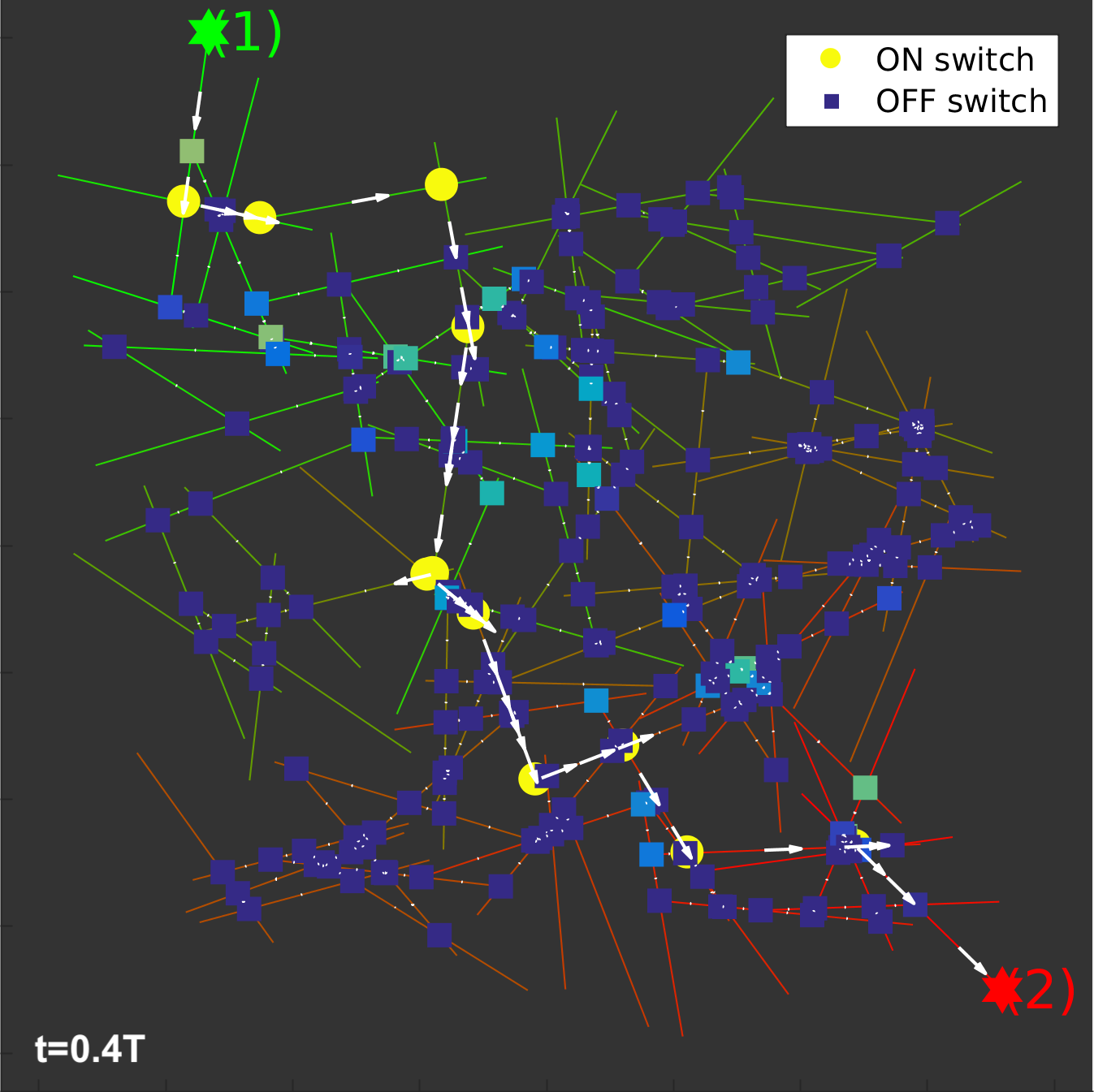}
    \includegraphics[width=1.7in]{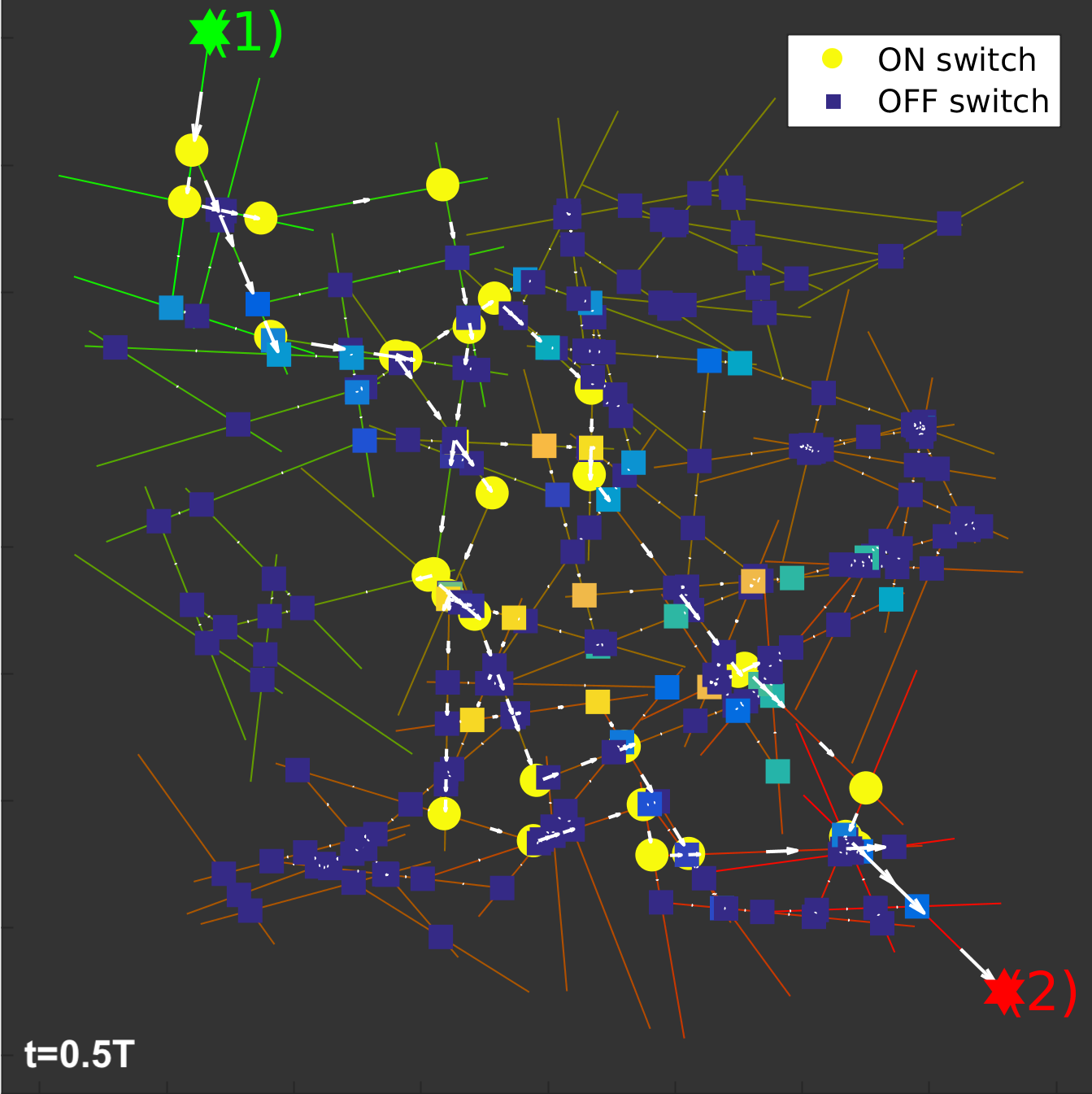}
    \includegraphics[width=1.7in]{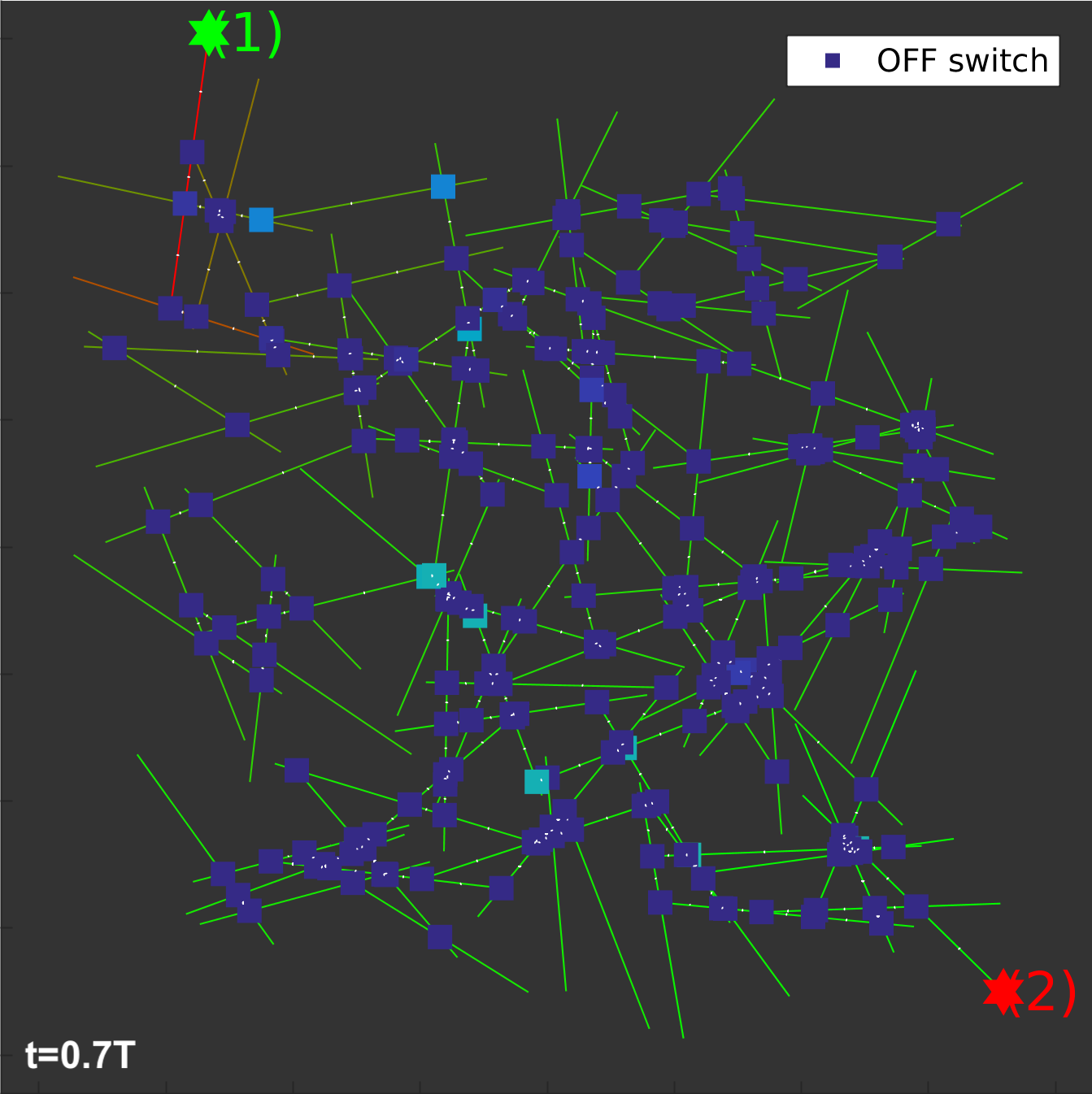}
    \caption{Top panel -- Individual memristive junction conductances $G_{\rm jn}$ (in units of conductance quanta $G_0$) as a function of time $t$ (in units of total simulation time $T$) for a triangular voltage signal (black) input to  a 261-junction NWN. Bottom panel -- snapshots of the network at four sequential time points ($t/T = 0.1, 0.4, 0.5, 0.7$) with colorbar indicating $G_{\rm jn}$. Dark blue junctions denote memristive switches in the off state.}
    \label{fig:adaptive}
\end{figure}

Figure~\ref{fig:adaptive} shows the NWN response to a triangular input signal. In the top panel, each colored curve represents the evolution in time of conductance $G_{\rm jn}$ across an individual memristive junction. The network connectivity determines the spatial distribution of voltage at each moment in time. This connectivity influences the voltage-controled memristive dynamics of each junction, resulting in collective switching as $G_{\rm jn}$ continuously adapts. The bottom panel shows this self-regulation of the synaptic junctions in snapshot visualizations of the network at successive time points during evolution. Brightly colored circles evident in the frames at $t = 0.4T$ and $t=0.5T$ represent memristive switches in their on state, with current paths indicated (white). The intrinsic adaptive dynamics of NWNs can in principle be harnessed for information processing. For the parameters used in Fig.~\ref{fig:adaptive} ($f = 0.75\,$Hz. $A = 0.8\,$V), the network exhibits ``edge-of'chaos'' dynamics (e.g. $I-V$ trajectories begin to diverge), which may be optimal for information processing \cite{edge-chaos}.     

\subsection{Mackey--Glass time--series prediction}

\begin{figure}[h]
    \centering
    \includegraphics[width=3.5in]{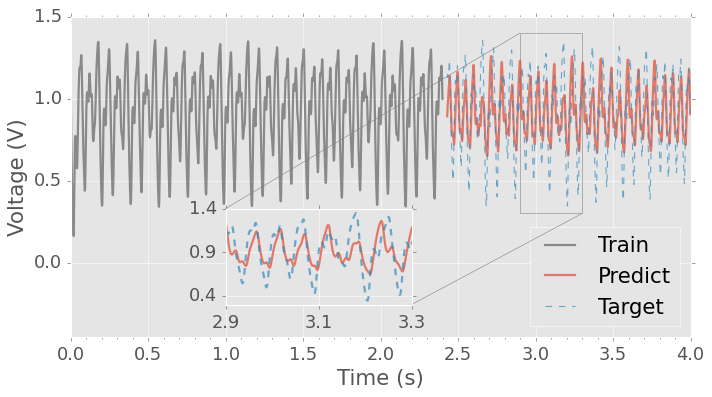}
    \caption{Time series of MG source signal during training (gray), followed by target (dashed blue) and predicted (red) signals for $\tau = 20$. The inset shows a zoom-in of part of the prediction period.}
    \label{fig:MG-prediction}
\end{figure}

\begin{figure}[h]
    \centering
    \includegraphics[width=3.0in]{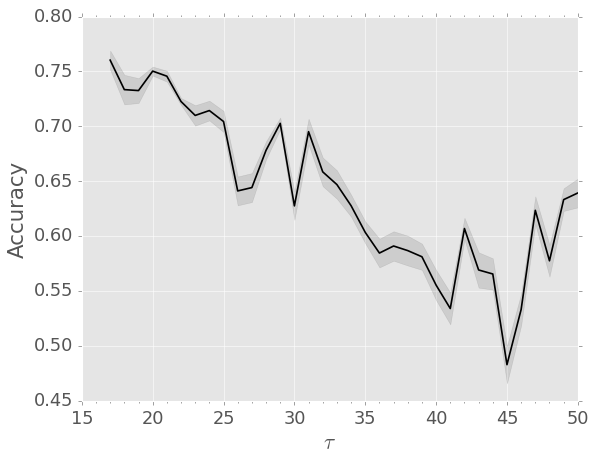}
    \caption{Average MG forecasting accuracy as a function of delay parameter $\tau$. Shading indicates standard error.}
    \label{fig:MG-tau}
\end{figure}

Fig.~\ref{fig:MG-prediction} plots the time series for training and predicting a MG signal with $\tau = 20$. Network output weights $\wvec$ are trained using the first 2400 time steps (i.e. $T = 2.4\,$s) after which the signal is predicted 20 steps ahead using eq.~(\ref{e:predict}) (i.e. with $\delta t =0.02\,$s in this case). The target signal is overplotted for comparison. The resulting prediction accuracy is 75\%.  
Fig.~\ref{fig:MG-tau} plots the prediction accuracy as a function of $\tau \geq 17$.
Accuracy decreases with $\tau$ because errors amplify exponentially as the MG signal becomes more chaotic.

\subsection{Transfer learning}

\begin{figure}[h]
    \centering
    \includegraphics[width=3.4in]{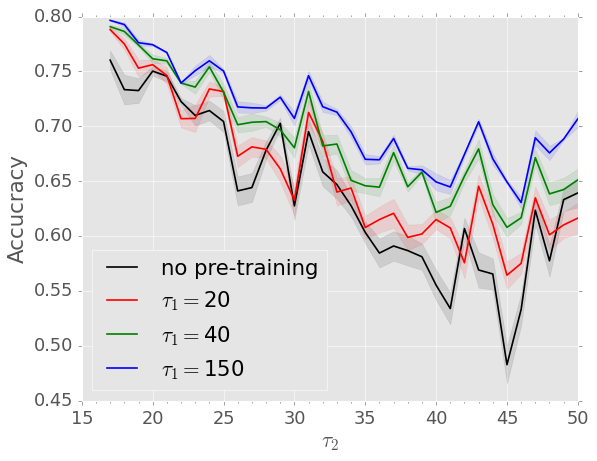}
    \caption{Average MG signal prediction accuracy as a function of signal delay parameter $\tau_2$ of the predicted signal and different $\tau_1$ signals used to prime the network before training. Average accuracy without pre-training is shown for comparison (cf. Fig.~\ref{fig:MG-tau}). Shading indicates standard error.}
    \label{fig:MG-transfer}
\end{figure}

Figure~\ref{fig:MG-transfer} plots MG prediction accuracy when the network is primed by a MG signal prior to training. Accuracy is plotted for three different pre-training MG signals ($\tau_1 = 20, 40, 150$) as a function of $\tau_2$ used for MG signal training. For comparison, prediction accuracy without the pre-training (cf. Fig.~\ref{fig:MG-tau}) is also overplotted.
Accuracy improves when the network is first primed with a MG signal. This demonstrates the principle of transfer learning, where knowledge is extracted from a source domain and then leveraged for learning in a related target domain.

\begin{figure}[h]
    \centering
    \includegraphics[width=3.4in]{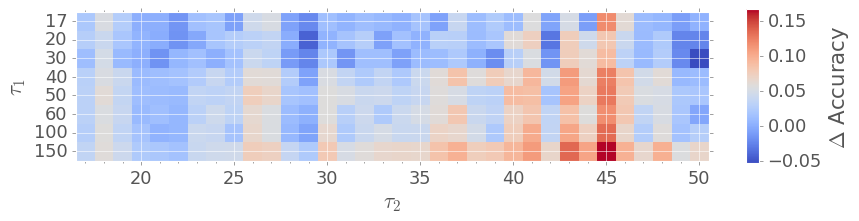}
    \caption{Heatmap showing change in average accuracy in predicting a MG signal with $\tau_2$ when the network is primed using a MG signal with $\tau_1$ relative to prediction without priming.}
    \label{fig:MG-transfer-diff}
\end{figure}

Learning performance is expected to improve especially when there is insufficient information in the target domain compared to the source domain. In this example, prediction accuracy improves more when the network is primed by a source MG signal that is more chaotic (i.e. more degrees of freedom) than the target MG signal (i.e. $\tau_1 > \tau_2$). This is shown by the blue curve (for $\tau_1 = 150$) in Fig.~\ref{fig:MG-transfer} and by the accuracy difference heatmap in Fig.~\ref{fig:MG-transfer-diff}.

Importantly, the target MG signal is predicted without relying on any teacher signal for recall. This suggests learning is achieved by harnessing the network's collective memory of past dynamical states. Priming the network before training improves learning by strengthening the memristive connections in an adaptive way, enabling longer-term memory consolidation.

Prediction accuracy also depends on the instantaneous network state selected for priming. Regardless of the value of $\tau_2$ of the MG signal being predicted, we find accuracy is optimized for a small range of primed network states. This optimal range of states occurs around network activation, coinciding with the formation of a winner-takes-all (WTA) current path (cf. Fig.~\ref{fig:adaptive}, bottom panel). Such WTA gate modules in network circuits are purported to have universal computational power for both digital and analog information processing \cite{Maass-2000}.

\section{Conclusions}

We have demonstrated that the complex interplay between the neural network-like circuitry of nanowire networks and their memristive junctions results in adaptive dynamics, where the network self-regulates to find the optimal signal transduction routes.
We showed how the adaptive dynamics can be harnessed for signal processing using a reservoir computing implementation. 
Prediction of the highly nonlinear Mackey--Glass signal was demonstrated well into the strongly chaotic regime. This has not previously been demonstrated with other memristive reservoir computing approaches.
Moreover, we found performance accuracy of this task is improved by transfer learning, where the network is primed by a Mackey--Glass signal before training. Our results show that transfer learning improves performance the most when pre-training with a source signal that is more complex than the target signal to be predicted. 

\section*{Acknowledgment}

The authors acknowledge use of the Artemis High Performance Computing resource at the Sydney Informatics Hub, a Core Research Facility of the University of Sydney.

\clearpage


\end{document}